\documentclass[10pt]{elsarticle}
\usepackage[utf8]{inputenc}
\usepackage{enumitem}  
\usepackage[english]{babel}
\usepackage{tcolorbox}
\usepackage{pgffor}
\usepackage{times}
\usepackage{silence}
\usepackage{varioref}
\usepackage{rotating}
\usepackage{wrapfig}
\usepackage{amssymb}
\usepackage{balance}
\usepackage{pifont}

\usepackage{array}
\newcolumntype{P}[1]{>{\centering\arraybackslash}p{#1}}

\usepackage{color, colortbl}
\definecolor{gray}{gray}{0.9}

\usepackage[scaled]{beramono}
\usepackage[T1]{fontenc}
\usepackage{array}
\newcolumntype{T}{l<{\verb+}}

\newcommand{\cmark}{\ding{51}}%
\newcommand{\xmark}{\ding{55}}%

\newcommand{\bi}{\begin{itemize}}
\newcommand{\ei}{\end{itemize}}

\usepackage{hyperref} 
\hypersetup{
    colorlinks=true,
    linkcolor=blue,
    filecolor=black,   citecolor=black,
    urlcolor=blue,
    pdftitle={Bad Smells in  Software Analytics Articles, Menzies \& Shepperd},
    bookmarks=true 
}

\newcommand{\RED}{\color{black}}
\newcommand{\RRED}{\color{black}}
\newcommand{\BLACK}{\color{black}}

\journal{Information \& Software Technology}

\begin{document}

\title{``Bad Smells" in Software Analytics Papers}
\author{Tim Menzies}
\address{Dept.\ of Computer Science\\North Carolina State University, USA}
\author{Martin Shepperd}
\address{Brunel Software Engineering Lab (BSEL)\\Dept.\ of Computer Science\\Brunel University London\\ UB8 3PH, UK}

\begin{abstract}
CONTEXT:  There has been a rapid growth in the use of data analytics to underpin evidence-based software engineering.  However the combination of complex techniques, diverse reporting standards and poorly understood underlying phenomena are causing some concern as to the reliability of studies.\newline
OBJECTIVE: Our goal is to provide guidance for producers and consumers of software analytics studies (computational experiments and correlation studies).\newline
METHOD: We propose using ``bad smells", i.e., surface indications of deeper problems and popular in the agile software community and consider how they may be manifest in software analytics studies.\newline
RESULTS: We  list 12 ``bad smells" in software analytics papers (and show their impact by examples).\newline 
CONCLUSIONS: \RRED We believe the metaphor of bad smell is a useful device. Therefore we encourage more debate on what contributes to the validty of software analytics studies \BLACK (so we expect our list will mature over time).
\end{abstract}


\maketitle

\section{Introduction}

\noindent
The drive to establish software engineering as an evidence-based discipline has been gaining momentum since the seminal article of Kitchenham et al.~\cite{Kitc04}.  In parallel there has been a massive growth in the amount of publicly available software data and sophisticated data analytics methods.  This has resulted in a sharp increase in the number and reach of empirically based, data driven studies that are generally termed software analytics.

Typically software analytics studies seek to distill large amounts of low-value data into small chunks of very high-value information. Studies are more data-driven than theory-driven.  For example, after examining many software projects, certain coding styles could be seen to be more bug prone.  Follow up experimentation, by manipulating the coding styles, could lead us to believe there is causality. Hence we might recommend some coding styles should be avoided (subject to various context-related caveats). \RED However, absence of theory can lead to challenges.  In particular, the lack of strong prior beliefs may make it difficult to choose between or evaluate potentially millions of possibilities \cite{Gust93,Deva16}.  \BLACK




\begin{table}
\caption{Software Analytics Bad Smells Summary}

\resizebox{!}{.5\textheight}{%
\scriptsize
\begin{tabular}{|r|p{1.2cm}|p{3cm}|p{2.7cm}|p{3.5cm}|}
\hline
\rowcolor{black}  & \textbf{\textcolor{white}{``Bad smell"}} & \textbf{\textcolor{white}{Core problem}} & \textbf{\textcolor{white}{Impact}} & \textbf{\textcolor{white}{Remedy}} \\
\hline
1 & Not interesting & The software analytics equivalent of how ``many angels can dance on a pinhead''. \RRED This may result from an absence of theory. \BLACK& Research that has negligible software engineering impact & Dialogue with practitioners \cite{Bege14,Lo2015,Ivar11} \\
\rowcolor{blue!10}2 & Not using related work & Unawareness of related work on (i) the research question and/or (ii) state of the art in relevant techniques. & (i) Unwitting duplication (ii) Not using state of the art methods (iii) Using unchallenging / outdated benchmarks & Read! Utilize systematic reviews whenever possible. \RRED Search for relevant theory or theory fragments. \BLACK Utilize technical guidance on data analytics methods and statistical analysis. Speak to experts. \\
3 & Using deprecated or suspect data e.g., $D$ has been widely used in the past ... & Using data for convenience rather than for relevance & Data driven analysis is undermined \cite{agrawal18} & Justify choice of data sets wrt the research question. \\
\rowcolor{blue!10}4 & Inadequate
\newline reporting & Partial reporting of results e.g., only giving means without measures of dispersion and/or incomplete description of algorithms & (i) Study is not reproducible \cite{Made17} (ii) meta-analysis is difficult or impossible. & Provide scripts / code as well as data. Provide raw results \cite{Good16,Muna17}. \\
5 & Under-powered studies & Small effect sizes and little or no underlying theory & Under-powered studies lead to over-estimation of effect sizes and replication problems & Analyse power in advance. Be wary of searching for small effect sizes, avoid Harking and p-hacking \cite{Butt13,Jorg16,Muna17}.\\
\rowcolor{blue!10}6 & $p<0.05$ and all that! & Over-reliance on, or abuse of, null hypothesis significance testing & A focus on statistical significance rather than effect sizes leading to vulnerability to questionable research practices and selective reporting \cite{Jorg16} & Report effect sizes and confidence limits \cite{Elli10,Lake14}. \\
7  & Assumptions of normality and equal variances in statistical analysis & Impact of heteroscedasticity and outliers e.g., heavy tails ignored.  Assumptions of analysis ignored.  & Under-powered studies lead to over-estimation of effect sizes and replication problems. & Use robust statistics / involve statistical experts \cite{Kitc02,Wilc12}.\\
\rowcolor{blue!10}8 & No data \newline visualisation & Simple summary statistics e.g., means and medians may mask underlying problems or unusual patterns. & Data are misinterpreted and anomalies missed. & Employ simple visualisations for the benefit of the analyst and the reader \cite{Mya09,Heal18}.\\
9 & Not exploring stability.  Only reporting best or averaged results. & No sensitivity analysis. p-hacking & The impact of small measurement errors and small changes to the input data are unknown but potentially substantial.  This is particularly acute for complex models and analyses. & Undertake sensitivity analysis or bootstraps/randomisation to understand variability. Minimally report all results and \textit{variances} \cite{Manl97,Salt00}.\\
\rowcolor{blue!10}
10 & Not tuning & Biased comparisons e.g., some models / learners tuned and others not. Non-reporting of the parameter settings, the tuning effort / expertise required. & Under-estimation of the performance of un-tuned predictors. Inability to reproduce results. & Read and use technical guidance on data analytics methods and statistical analysis \cite{Snoe12,fu2016}.  Unless relevant to the research question avoid off-the shelf defaults for sophisticated learners.\\
11 & Not exploring simplicity & Excessively complex models, particularly with respect to sample size  & Dangers of over-fitting & Independent validation sets or uncontaminated cross-validation \cite{Cawl10}. \\
\rowcolor{blue!10} 
  12 & Not justifying choice of learner & Permuting / recombining learners to make `new' learners in a quest for spurious novelty & Under / non reporting of ``uninteresting results" leading to over-estimates of effect sizes. & Principled generation of research questions and analysis techniques. Full reporting. \\
\hline

\end{tabular} 
}
\label{Tab:BadSmells}
\end{table}%

Thus far, so good.  Unfortunately, concerns are being expressed about the reliability of many of these results both from within software engineering (SE) \cite{Shep14,Jorg16} and more widely from other experimental and data driven disciplines such as the bio-medical sciences \cite{Ioan05,Earp15}.  This has reached the point that in experimental psychology researchers now refer to the ``replication crisis''.  This in turn raises questions of study validity.

Arguably the situation is not  dissimilar in software engineering.  In a major systematic review of SE experiments (1993--2002) Kampenes et al.~\cite{Kamp07} found similar effect sizes being reported to psychology. In 2012 we published an editorial for a Special Issue in the journal \textit{Empirical Software Engineering} on repeatable results in software engineering prediction \cite{menzies12} where the principal focus was ``conclusion instability".  In that editorial, we raised concerns about the numerous occasions where Study 1 concluded X yet Study~2 concluded $\neg$X.  The inability to resolve these seeming contradictions leaves us at an impasse.  Clearly, we need better ways to conduct and report our work, as well as to examine the work of others.

In this article we consider what may be learned from other disciplines and what specific lessons should be applied to improving the reporting of software analytics type studies.  ``Bad smells'' is a term that comes from the agile community. According to Fowler~\cite{beck1999bad}, bad smells (a.k.a.\ code smells) are ``a surface indication that usually corresponds to a deeper problem''.  See Table~\ref{Tab:BadSmells} for a summary of the bad smells identified in this article.

As per Fowler, we say that the ``smells''  listed in this article do not necessarily indicate that the conclusions from the underlying study must be rejected.  However, we believe they raise three areas of potential concern.
\begin{enumerate}
    \item For the author(s) they reduce the likelihood of publication.
    \item For the reader they raise red flags concerning the reliability of the results \RED (i.e., their consistency) and the validity (i.e., the correctness of the analysis or how well it captures those constructs we believe it represents) \BLACK.
    \item For science, they hinder the opportunities of pooling results via meta-analysis and building bodies of knowledge. 
\end{enumerate}
So the goal of this article is to identify ``smells'' in order to (i) encourage a wide-ranging community debate and (ii) assist in the identification of potential problems and associated remedies.  

The remainder of this article is organized as follows. The rest of this section addresses scope and then tackles two frequently asked questions about this work. Section \ref{tion:RelWk} explores historical work on writing `good' articles and methodological guidance, coupled with how other data-driven disciplines have addressed experimental reliability concerns.  Next, in Section \ref{tion:Approach} we describe our approach to identifying ``bad smells''.  This is followed by Section \ref{tion:BadSmells} that lists some key symptoms or `smells' in approximate order of importance.  Finally, in Section \ref{tion:Disc} we consider the implications for the future and how we, the research community, might collectively take ownership of these ``smells'', re-prioritise, re-order, add and subtract ``smells''.

\subsection{Scope}
Software analytics research is relatively new and growing fast, so much of our advice relates to articles discussing induction from software project data. We discuss problems with empirical data analysis via experiments, quasi-experiments and correlation studies \cite{Shad02} rather than issues relating to e.g., qualitative studies.  Other articles should be consulted for tips and traps relating to qualitative research and experimentation involving human participants, e.g., \cite{stol2016grounded,cruzes2011research,ko2015practical,petersen2015guidelines}.

\RRED
Finally, in order to delineate the scope of this article we feel it helpful to add some remarks concerning the role of theory in software analytics type research.  Naturally much analytics research will be data-driven or inductive in nature.  In such circumstances the role of theory or deductive reasoning will be secondary.  This of course begs the question of what is a theory.  Unfortunately there is limited agreement as to what constitutes a theory \cite{Sjob08,John12}, other than it may comprise constructs, relationships between constructs (possibly causal ones), scope and possibly propositions \cite{Stol13}. Most commentators are of the view that theory use is low and might fruitfully be increased.  As a step towards this Stol and Fitzgerald \cite{Stol13} explore the notion of theory fragments which might take the form of propositions.  Those that are testable might form hypotheses i.e., a mechanism from theory to the empirical world and putative propositions as a means of navigating from empirical observations back to theory.

Having said this, we do not include a detailed treatment of software engineering theory.  Data analytics studies are almost always theory light because they're  inductive in their approach.  The use of theory, or not, will be highly context sensitive.  There is not presently consensus as to what theories (or for that models) should look like.  Finally, as indicated, there are a number of good treatments that already exist (see \cite{Sjob08,John12,Stol13}).
\BLACK

\subsection{But is this article necessary?}
Simply put: is this article necessary? Is not everything here just material that ``everybody knows''?

In response, we first note that, over some decades, the authors have been members of many conference and journal review panels. Based on that experience, we assert that demonstrably everybody does not know (or at least apply) this material.  Each year, numerous research articles are rejected for violating the recommendations of this article, or worse still, some slip through the net.  In our view many of those rejections are avoidable. We believe, many of the issues raised by this article can be repaired before submission -- some with very little effort.

These difficulties are compounded by the fact that good practice evolves.  For example, whilst the statistical education of many engaged in empirical software engineering research is grounded in classical statistics there have been more recent developments such as robust approaches \cite{Kitc17}, use of randomisation and bootstrap \cite{Carp00} and Bayesian techniques \cite{Gelm13}.  Likewise, there are significant and recent developments in machine learning which, sometimes, authors overlook. For example, we have had on occasions to review and reject papers that are some very small delta to an article one of us wrote in 2007~\cite{menzies07}. Since that paper was written, much has changed\footnote{For example, improved evaluation techniques and experimental design; new problem areas for SE and machine learning (e.g.~\cite{sarkar2015cost});
 novel and insightful new evaluation measures (see Section 5 of~\cite{huang2017supervised} or the multi-objective approaches discussed in~\cite{nair2018data}; and challenging new learners that are so slow that it becomes impractical to reproduce prior results (see the discussion on deep learning in~\cite{fu2017easy}).} so a mere increment on that 2007 paper is unlikely to be a useful contribution or an acceptable publication. 

We also observe that, on occasions, we find some reviewers who have the role of gatekeepers to scientific publication, also seem unaware of some, or much of this material.  Thus there exists some research that should not be published, at least in its present form, that is published \cite{Kitc02,Jorg16}.  This harms the progress of science and our ability to construct useful bodies of knowledge relevant to software engineering.
 
A particularly marginalized group are industrial practitioners trying to get their voice heard in the research literature.  Unless that group knows the established norms of publication, their articles run the risk of rejection. We note that those norms of publication are rarely recorded -- a problem that this article is attempting to solve.  Hence, for all the above reasons, we argue that this article is necessary and useful.

\RED \subsection{Should papers be rejected if they have bad smells?}\label{tion:reject}

It is important to stress that this paper:
\bi
\item
is primarily intended as \textit{guidance} for authors;
\item
is {\bf not} intended to be used by reviewers as a mindless checklist of potential reasons-to-reject a paper.  
Our bad smalls are {\em indicative} rather than {\em definitive} proof of a potential issue.  Hence we stress that {\em their presence alone is not a reason for reviewers to reject a paper}.   Context is extremely important.
\ei
One of the reviewers of this paper was kind enough to expand on this point. They commented (and we agree) that reviewers tend to become overly strict --- to the point where they can reject papers with very interesting and important ideas --- just because there was one point that did not satisfy certain guidelines.  This is most acute for conferences where there is seldom a revision cycle. Reviewers should not just blindly enforce the adoption of all the points raised by this paper, but also need to be \textit{sensitive} to the context of the study; certain bad smells may not be applicable (or may be less relevant).   Additionally, we wish to avoid adopting a perspective, such that the requirement of a perfection prevents useful, pioneering or relevant --- but slightly imperfect --- research from ever being published.
\BLACK

\section{Related Work}\label{tion:RelWk}
 
Software engineering (SE) researchers have posed the question from time to time: what makes a good article?  Examples are the article by Shaw \cite{Shaw03} back in 2003 and recently revisited by Theisen et al.~\cite{Thei17}.  Shaw stated that one difficulty SE researchers and authors face is that there are no well-developed or agreed guidelines as to how research should be conducted and presented.  Although in the intervening years there has been increasing emphasis placed upon evidence, it is our view that the situation has not greatly changed.  She also remarked on the weakness of many articles in terms of  validation and Theisen et al.\ report that, if anything, reviewer expectations have increased.  Potentially, software analytics articles have an advantage here given their empirical basis, however statistical analysis is not always undertaken well, as our article will demonstrate.

There have also been various audits of statistical quality of empirical research \cite{Kitc02} and published guidelines and recommendations regarding analysis \cite{Kitc02,Wohl12}. The guidelines are welcome since, by and large, software engineers have limited formal training as statisticians.  Note that when researchers do publish methodological guides, those articles tend to have very high citation counts\footnote{The citation counts were extracted from Google Scholar, November 2018.}, for instance:
\bi
\item ``Preliminary guidelines for empirical research in software engineering", in \textit{IEEE Transactions on Software Engineering} \cite{Kitc02} (1430 citations since 2002).
\item ``Benchmarking Classification Models for Software Defect Prediction: A Proposed Framework and Novel Findings'' in \textit{IEEE TSE} 2008~\cite{Lessmann08} (765 citations since 2008).
\item
``Guidelines for conducting and reporting case study research in software engineering'', 
  in the \textit{Empirical Software Engineering} journal~\cite{runeson2009guidelines} (2500 citations since 2009).
\item ``A practical guide for using statistical tests to assess randomized algorithms in software engineering'' at ICSE'11~\cite{Arcuri:2011} (500 citations since 2011).
\ei

Of particular concern is our our ability to replicate results (or reliabiity), which many argue is the cornerstone of science.  This was investigated systematically in experimental psychology by Nosek et al.~\cite{Open15} who concluded from a major replication project of 100 experiments in leading psychology journals that ``replication effects were half the magnitude of original effects'' and that while 97\% of the original studies had statistically significant results, only 36\% of replications did.  Researchers began to talk about the ``replication crisis''.  Note, however, that there has been little discussion as to exactly what constitutes a confirmation of the original study \cite{Spen16} and that for under-powered studies (as seems to be typical) results may well be dominated by sampling error.  Nevertheless, it seems that there is legitimate cause for concern and this has triggered recent how-to type articles such as Munaf{\`o} et al.~\cite{Muna17}.

Despite this ``replication crisis'' having sparked considerable soul searching when it was reported, less progress is being made than we would hope e.g., in cognitive neuro-science and experimental psychology study power remains low and essentially unchanged over 50 years \cite{Szuc17}.  Smaldino and McElreath \cite{Smal16} make the case that given the incentive structures for researchers --- specifically that publication is the principal factor for career advancement --- it is unsurprising that low quality research studies are still being conducted, submitted and published.  As they put it, although this occurs without ``deliberate cheating nor loafing" it helps explain why improvements in scientific practice remain slow.  

We believe that whilst incentive structures clearly matter and influence scientific behavior, many of the problems we identify could be remedied with relatively little effort.  We also hope that the intrinsic motivation of ``doing good science" will be sufficient for the large majority of our research community.

\section{Our Approach}\label{tion:Approach}

Our goal is to assist researchers towards conducting better research and writing better articles.  As steps towards that goal, we have a two-part proposal:
 
\begin{enumerate}
    \item A part of this proposal is to run workshops at major software engineering conferences where best, and worst, practices in analysis and writing can be explored and documented.  Accordingly, this article aims to recruit colleagues to this goal and thereby increase the number of  researchers reflecting and commenting on these issues.
\item The other part of this plan is this article itself.  Whilst we hope that these ideas will prove useful in their own right, we also hope that (a)~this article inspires more extensive discussions about best and worst paper writing practices; and (b)~subsequent articles extend and improve the guidance we offer below.
\end{enumerate}

In order for this article to be effective we adopt a  pragmatic approach.  We present, a hopefully useful but not exhaustive, list of problems with software analytics papers. The level of detail is illustrative rather than in depth.  In the spirit of brevity we restrict our focus to data driven research so, for example, case studies and experiments involving human participants are excluded.

\begin{table}[!t]
\caption{Cross referencing research area to bad smell }
\small
\begin{center}
\begin{tabular}{|p{3cm}|p{6cm}|p{2cm}|}
\hline
\textbf{Research area} \cite{Kitc15} & \textbf{Quality instrument question} \cite[Table 3]{Dyba08} & \textbf{Bad smell}\\
\hline
\textit{Reporting} - quality  & Is the article based on research? & Yes, defined by our scope \\
& Is there a clear statement of the research aims? & 1 \\
& Is there an adequate description of the context? & 4 \\
& Is there a clear statement of findings? & 4, 11 \\
\hline
\textit{Rigour} -  details of  & Was the research design appropriate? & 5 \\
the research design & Was the recruitment strategy appropriate? & No human participants, but more generally: 1, 3 \\
& Treatments compared to a  control group? & 2, 10 \\
& Data collected   addressed   research issue? & 3 \\
& Was the data analysis sufficiently rigorous? & 6, 7, 8, 9, 10 \\ 
\hline
\textit{Credibility} - are the findings of the study are valid and meaningful? & Has the relationship between researcher and participants been adequately considered?& No human\newline participants \\
\hline
\textit{Relevance} -  of the study to practice & Is the study of value for research or practice? & 1, 11\\
\hline
\end{tabular}
\end{center}
\label{Tab:Xref}
\end{table}%

Some ``smells" are more self-evident and consequently need relatively little explanation.  In contrast, others are more subtle and therefore we devote extra space to illustrating them.  We use examples to help illustrate the smells and their relationship to actual problems.  We do not intend examples as existence proofs, and in any case consider it invidious to single out individual authors as examples of widespread errors.

The list of ``smells" has been constructed by discussion and debate between the two authors.  We then cross-checked the list (see Table \ref{Tab:Xref}) with an empirical study quality instrument, \cite{Dyba08} that has been used when assessing article quality for inclusion within software engineering systematic reviews \cite[chapter 7]{Kitc15}.  Hence we see our list addresses all four areas of quality identified by the quality instrument and have more confidence in its coverage.

\section{``Bad Smells'' Unpacked}\label{tion:BadSmells}
This section discusses the bad smells listed in Table \ref{Tab:BadSmells}.
Before beginning, we note that while some of the following ``smells'' are specific to software analytics, others are general to many scientific endeavors.  We include them altogether since they are all important issues about good science and effective scientific communication.

\subsection{Not Interesting}

All science studies should strive to be valid and reliable. But it is important that research is also interesting (to a wider audience than just the authors).  This is not a clarion call for quirky or quixotic research, however, recall that our discipline is software \textit{engineering}. Potentially somebody must be invested in the answers, or potential answers.  One rule-of-thumb for increasing the probability that an article will be selected is validity plus {\em engagement} with some potential audience. 


So what do we mean by ``interesting''?  The article needs a motivation or a message that beyond mundane.  More specifically, {\em articles should have a point}. Research articles should contribute to a current debate. \RRED This could be via theory or with respect to the results of other studies. \BLACK If no such debate exists, research articles should motivate one.  For example, one could inspect the current literature and report a gap in the research that requires investigation.  But to be clear, this does not preclude incremental research or even checking the reproducibility of results from other researchers, particularly if those results may have strong practical significance.

Note that problem of {\em not inspiring} is particularly acute in this era of massive availability of data and new machine learning algorithms; not all possible research directions are actually interesting.  The risk can be particularly high when the work is data-driven which can quickly resemble data-dredging. For more on this point, see our last two bad smells ({\em Not enough theory}and {\em Not much theory}).

An antidote and powerful way to consider practical significance is with effect size \cite{Elli10} especially if this can be couched in terms that are meaningful to the target audience.  For instance, reducing unit test costs by $x\%$ or $y\$ $ is inherently more engaging than the associated $p$ value of some statistical test is below an arbitrary threshold. The notion of smallest effect size of interest has also been proposed as a mechanism to maintain focus on interesting empirical results \cite{Lake17} a point we shall return to.

\subsection{Not Using Related Work}
This issue should not exist. Nevertheless, we know of many articles that fail to make adequate use of related work. Typically we see the following specific sub-classes of ``bad smell''.

 {\em The literature review contains no or almost no articles less than ten years old.}  \RED Naturally older papers can be very important-- but the absence of any recent contribution would generally be surprising. \BLACK This strongly suggests the research is not state of the art, or is not being compared with state of the art.

{\em The review overlooks relevant, recent systematic literature review(s).}  This strongly suggests that there will be gaps or misconceptions in the way the state of the art is represented.  It may also allow bias to creep in since the authors' previous work may become unduly influential.

{\em The review is superficial, even perfunctory}. Such an approach might be caricatured along the lines of: \begin{quote}
    ``Here is some related work [1, ..., 30].  Now to get on with the important task of my describing own work."
\end{quote}
This style of writing --- where reviewing the work of others is merely a burden --- makes it less likely connections will be seen, that the article contributes to a body of knowledge but conversely, increases vulnerability to confirmation biases \cite{Nick98}.  It is also unhelpful to the reader since there is little or no interpretation of other work.  What is useful?  What has been superseded?

{\em The review is thorough, relevant and fair, however, it is not used to motivate the new study. } In other words there is a disconnect between the review  and new research. New results are not related to the current state of the art.  The consequence is the reader cannot determine what new contributions have been made.  It may also mean appropriate benchmarks are missed.

{\em Methodological aspects of the research are motivated by extremely simplistic or outdated sources e.g., using traditional statistics texts. }  One of the many dangers from using such sources is to buttress obsolete or inappropriate data analysis techniques or learning algorithms that are no longer competitive.

\begin{wrapfigure}{r}{1.8in}
  \centering
 \includegraphics[width=1.8in]{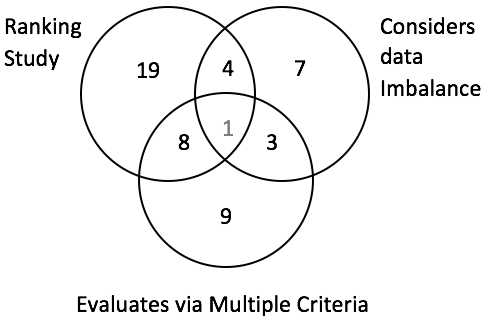}
  \caption{Summary of Table~\ref{tbl:survey2}.}
\label{fig:s2}
\vspace{-0.4cm}
\end{wrapfigure}
One constructive suggestion, especially if no relevant systematic literature review exists, is to employ {\em lightweight literature inspections}.  These need not be as complex as a full systematic literature review (that can take months to complete). 
However, with the growing number of published systematic literature reviews it is increasingly probable that other researchers will have already addressed at least some part of the literature relevant to the research topic you are exploring.

\begin{table}[!t]
\centering
\scriptsize
\caption{21 highly-cited software defect prediction studies,
selected as follows: (a)~from Google Scholar, search for
``software''
and ``defect prediction''
and ``OO'' and ``CK'' in the last decade;
(b)~delete anything with
less than ten cites/year since publication;
(c)~search the resulting
21 articles for the three research items explored by 
Agrawal et al., i.e., do they rank multiple classifiers;
are the classifiers ranked by multiple criteria;
does the study consider class imbalance.
}
\vspace{-0.2cm}
\label{tbl:survey2}
    \begin{tabular}{c|c|c|c|c|c}
        \begin{tabular}[c]{@{}c@{}}\textbf{Reference}\end{tabular} & \textbf{Year} & \textbf{Cites} & \begin{tabular}[c]{@{}c@{}}\textbf{Ranked} \\\textbf{Classifiers?} \end{tabular} &\begin{tabular}[c]{@{}c@{}} \textbf{Evaluated} \\\textbf{using} \\\textbf{multiple} \\\textbf{criteria?}\end{tabular}&\begin{tabular}[c]{@{}c@{}} \textbf{Considered}\\\textbf{Data}\\\textbf{Imbalance?} \end{tabular}\\ \hline
       (a) & 2007 & 855 & \cmark & 2 & \xmark \\
         (b) & 2008 & 607 & \cmark & 1 & \xmark \\ 
         (c) & 2008 & 298 & \cmark & 2 & \xmark\\  
         (d) & 2010 & 178 & \cmark & 3 & \xmark \\  
         (e) & 2008 & 159 & \cmark & 1 & \xmark\\     
         (f)& 2011 & 153 &\cmark & 2 & \xmark \\ 
         (g) & 2013 & 150 & \cmark & 1 & \xmark \\   
         (h) & 2008 & 133 & \cmark & 1 & \xmark \\    
         (i)& 2013 & 115 & \cmark & 1 & \cmark \\  
       (j) & 2009 & 92 & \cmark & 1 & \xmark \\          
        (k) & 2012 & 79 & \cmark & 2 & \xmark  \\ 
         (l) & 2007 & 73 & \xmark & 2 & \cmark\\  
         (m)& 2007 & 66 & \xmark & 1 & \cmark \\  
       (n) & 2009 & 62 & \cmark & 3 & \xmark  \\ 
   (o)& 2010 & 60 & \cmark & 1 & \cmark  \\  
        (p) & 2015 & 53 & \cmark & 1 & \xmark  \\  
         (q) & 2008 & 41 & \cmark & 1 & \xmark  \\  
         (r) & 2016 & 31 & \cmark & 1 & \xmark  \\ 
       (s) & 2015 & 27 & \xmark & 2 & \cmark \\  
        (t) & 2012 & 23 & \xmark & 1 & \cmark \\  
        (u) & 2016 & 15 & \cmark & 1 & \xmark  
\end{tabular}
\vspace{-0.3cm}
\end{table}

For examples of lightweight literature inspections, see Figure~\ref{fig:s2} and Table~\ref{tbl:survey2} from Agrawal and Menzies~\cite{agrawal18}. Here, the researchers found a research gap in two dozen articles of the top-most cited software defect prediction studies that have at least 10 citations per year over the last decade. Specifically, as seen in Figure~\ref{fig:s2}, most articles avoid issues of data imbalance and scoring via multiple criteria.  \RED Of course, research gaps need to be interesting (see Bad Smell \#1). \BLACK This finding led Agrawal and Menzies to perform a unique set of experiments that resulted in an ICSE'18 article.

\subsection{Using Deprecated or Suspect Data}

Researchers adopting a heavily data-driven approach, as is the case for software analytics, need to be vigilant regarding data quality.  De Veaux and Hand point out ``[e]ven when the statistical procedure and motives are correct, bad data can produce results that have no validity at all'' \cite{DeVe05}.  A series of systematic reviews and mapping studies \cite{Lieb08,Bosu13,Rosl13,Lieb16} all point to problems of data quality not being taken very seriously and problems with how we definitively establish the validity of data sets, particularly when the gap in time and perhaps geography, between the collection and analysis is wide.

In terms of ``smells" or warning signs we suggest researchers be concerned by the following:
\begin{itemize}
\item Very old data; e.g., the old PROMISE data from last-century NASA project (JM1, CM1, KC1, KC2, ...) or the original COCOMO data sets containing size and effort data from the 1970s.
\item Problematic data; e.g., data sets with known quality issues (which, again, includes the old PROMISE data from last-century NASA project~\cite{shepperd13}).
\item Data from trivial problems.  For example, in the history of software test research, the Siemens test suite
was an important early landmark in creating reproducible research problems. Inspired by that early work, other researchers have take to document other, more complex problems\footnote{e.g. See the Software-artifact Infrastructure Repository at
\href{http://sir.unl.edu/portal}{sir.unl.edu/portal}.}. Hence, a new paper basing all its conclusions on the triangle inequality problem from the Siemens suite is deprecated.   Nevertheless tiny problems occasionally offer certain benefits, e.g., for the purposes of explaining and debugging algorithms, prior to attempting scale up. 
\end{itemize}

That said, our focus is empirical studies in the field of software engineering where scale is generally the issue. In our experience, readers are increasingly interested in results from more realistic sources, rather than results from synthetic or toy problems. Given there now exist many on-line repositories where researchers can access a wide range of data from many projects (e.g. GitHub, GitTorrent\footnote{\href{https://github.com/cjb/GitTorrent}{github.com/cjb/GitTorrent}}; and others~\footnote{\href{http://tiny.cc/seacraft}{tiny.cc/seacraft}}) this should not be problematic. Consequently, it is now normal to publish using data extracted from dozens to hundreds of projects e.g., \cite{rahulSEIP, amritSeip18b}.

Finally, we acknowledge there are occasions when one must compromise, otherwise important and interesting software engineering questions may never be tackled.  Not all areas have large, well-curated, high quality data sets.  In such circumstances research might be seen as more exploratory or to serve some motivational role in the collection of better data.  As long as these issues are transparent we see no harm.

\subsection{Inadequate Reporting}\label{tion:report}

Here there are two dimensions: completeness and clarity. Problems may be indicated by the following ``smells".

\begin{enumerate}
\item The work cannot be reproduced because the descriptions are too incomplete or high level.  In addition, not all materials ( data and code) are available \cite{Good16,Muna17}.
\item The work cannot be \textit{exactly} reproduced because some of the algorithms are stochastic and seeds are not provided.  Typical examples are the generation of folds for cross-validation in machine learning and heuristic search algorithms.
\item Non-significant and post hoc ``uninteresting" results are not reported. This seemingly innocuous behavior, known as reporting bias, can lead to serious over-estimation of effect sizes and harm our ability to build bodies of knowledge via meta-analysis or similar means \cite{Good16,Szuc17}.
\item The experimental details are full of ``magic" numbers or arbitrary values for parameters and the experimental design. It is important to justify choices otherwise there are real dangers of researcher bias and over-fitting, i.e., choosing values that favor their pet algorithms.  Standard justifications include ``$k$ was selected via the standard Gap statistic procedure'' or perform some sensitivity analysis to explore the impact of choosing different values \cite{Salt00}.
\item It's impossible for the reader to get a sense of what the article is about without reading it in its entirety. Structured abstracts (as per this article) can greatly assist both human readers \cite{Budg11,4460893,booth1997value} and text mining algorithms which are beginning to be deployed to assist meta-analysis, e.g., \cite{Szuc17}.
\item Using highly idiosyncratic section headings or article organization.  Typically the following should suffice: introduction, motivation, related work, methods (which divides into data, algorithms, experimental method, and statistical procedures) followed by threats to discussion, threats to validity, conclusion, future work and references.  Simple tables summarising algorithm details, hyper-parameters etc can be invaluable. 
\item The article uses too many acronyms.  Define terms when they are first used.  If there are many, then consider using a glossary.  

\end{enumerate}
We conclude this sub-section with some simple and hopefully constructive suggestions.

{\em Strive towards reproducible results:}
Table~\ref{tbl:acm} defines a continuum of research artifacts and results.  Reporting reproducible results (on the right-hand-side of that continuum) is the ideal case -- but that can take a great deal of effort.  
For an excellent pragmatic discussion on how to structure your research artifacts in order to enhance reproducibility, see ``Good Enough Practices for Scientific Computing''  \href{http://goo.gl/TD7Cax}{goo.gl/TD7Cax}.

\RED(Aside: Demands for increased reproducibility must be assessed on pragmatic grounds.  If a prior study uses tools that are not open source, or tools that are very complex to use, then it is not appropriate to demand that subsequent work exactly reproduces the old work. )\label{noreprod} \BLACK

\begin{table}[!t]
\caption{The artifact continuum flows from left to right. More articles will classified on the right once authors apply more of the methods on the left.  This table presents the badges and definitions defined by the ACM (see \href{https://goo.gl/wVEZGxx}{goo.gl/wVEZGx}).
These definitions can be applied by the broader community as
(1)~aspirational goals, or (2)~ways to audit past research, or (3)~as guidance for organizing an ``artifacts track'' at  a conference or journal.}\label{tbl:acm}
\scriptsize

\vspace{5mm}
\begin{tabular}{c|P{0.6in}|P{0.8in}|P{0.9in}|P{0.6in}|P{0.6in}}
 No&\multicolumn{3}{|c|}{Artifacts} & \multicolumn{2}{c}{Results} \\\cline{2-6}
  badge & Functional & Resuable &Available & Replicated & Reproduced\\\hline

& \includegraphics[width=0.6in]{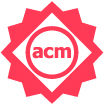}  

& \includegraphics[width=0.6in]{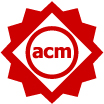}  

& \includegraphics[width=0.6in]{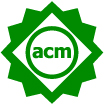}  

& \includegraphics[width=0.6in]{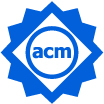}  

& \includegraphics[width=0.6in]{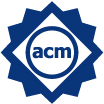}  \\
& Artifacts documented, consistent, complete, executable, and include appropriate evidence of verification and validation.
&
Functional + very carefully documented and well-structured to the extent that reuse and re-purposing is facilitated. In particular, norms and standards of the research community for artifacts of this type are strictly adhered to. &
Functional + placed on a publicly accessible archival repository. 

A DOI (Digital Object Identifier) or link to this repository along with a unique identifier for the object is provided. It is easy to obtain such DOIs: just host your artifact at repositories like SEACRAFT
\href{https://tiny.cc/seacraft}{tiny.cc/seacraft}. &
Available + main results of the article have been obtained in a subsequent study by a person or team other than the authors, using, in part, artifacts provided by the author.  &
Available + the main results of the article have been independently obtained in a subsequent study by a person or team other than the authors, without the use of author-supplied artifacts. \\ 
\end{tabular}
\end{table}

{\em Consider adding summary text to each table or figure caption}, e.g., ``Figure X: Statistical results showing that only a few learners succeed at this task'':  Quite often, reviewers will first skim  a article, glancing only at the figures. For such a reader, the locality of these annotations may prove extremely helpful.  

{\em Considering using research questions:} Frequently software analytics research articles naturally sub-divide themselves into exploring a small set of research questions. Determining these research questions can require some creativity but is a powerful structuring device. Preferably there should be some progression so for example, here are the research questions from~\cite{fu2016} (which discussed the merits of tuning the hyper-parameters of data mining algorithms):

\bi
\item RQ1: Does tuning improve the performance scores of a predictor? 
\item RQ2: Does tuning change conclusions on what learners are better than others? 
\item RQ3: Does tuning change conclusions about what factors are most important in software engineering? 
\item RQ4: Is tuning impractically slow? 
\item RQ5:
Is tuning easy?
\item RQ6: Should data miners be used ``off-the-shelf'' with their default tunings? 
\ei


A useful typographical device sometimes associated with research questions is the use of results boxes to stress the answers to the questions. Here is an early example from Zimmermann et al.~\cite{Zimmermann04}, 2004. Although potentially powerful such summaries need to be carefully constructed to avoid trivial over-simplification.

\centerline{\includegraphics[width=3.5in]{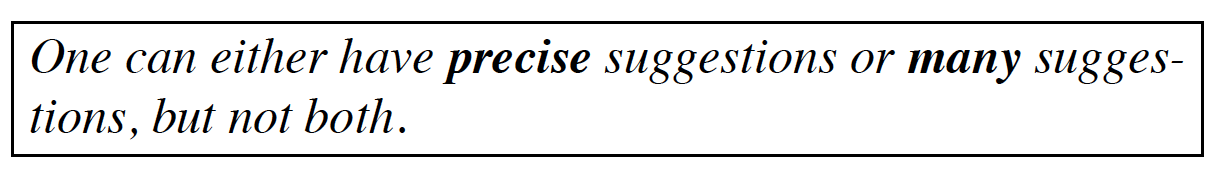}}


\subsection{Under-powered Studies}

The power of an analysis is the probability that a test will reject a false null hypothesis i.e., correctly report a non-zero effect. \RED More generally, it is the ability to find an effect were one to exist.  Power \BLACK is the complement of $\beta$ or the probability of making a Type II (failing to detect a true effect in favour of the null hypothesis, i.e., a false negative) error, hence power = $1 - \beta$.  
\RED However, there is a tradeoff with the probability of making a Type I error (i.e., a false positive).  At an extreme we could guard against false positives by setting an extremely high threshold of acceptance but the consequence would be an elevated false negative rate.  And of course \textit{vice versa}. \BLACK

\RED So what should $\beta$ be set to?  \BLACK Traditionally, it has been argued that power should be 0.8 \cite{Cohe88} indicating that researchers might consider Type I errors (wrongly rejecting the null hypothesis when in reality there is no effect) as being four times more problematic than Type II errors and therefore needing proportionately more checks. Thus, if significance levels are set at $\alpha=0.05$ then it follows that $\beta = 4\alpha = 0.2$ hence, as required, our power is $1 - \beta = 0.8$.  To repeat, the important point is that we must trade off between Type I and Type II errors and our preferences are somewhat arbitrary.

\RED However, what we want the power of a study to be, and what it \textit{actually} is are two different things.  
\BLACK 
Four factors influence power, namely: (i) sample size, (ii) the unknown, true size of the effect being investigated, (iii) the chosen $\alpha$ and (iv) the chosen $\beta$ \cite{Cohe92,Elli10}.  In addition, experimental design can improve power, for example by measuring within experimental unit (e.g., via a repeated measures design) rather than between unit variation \cite{McCl00}.  

Many commentators recommend, the power of a proposed experiment or investigation should be determined \emph{a priori} \cite{Dyba06}.  The purpose is to check if an experiment has a reasonable chance of detecting an effect were one to actually exist.  One obvious and immediate difficulty is estimating the effect size that is to be investigated \cite{Maxw08}.  This is compounded by the problem that there is compelling evidence that we are systematically over-estimating effect sizes due to reporting and publication bias \cite{Butt13,Jorg16}.  However, the problems of power go beyond merely wasting scientific resources in failing to detect effects that actually exist.  There are two other serious negative consequences.  First is the tendency of over-estimation of the measured effect sizes and second, is the elevated probability of false statistically significant findings \cite{Ioan05,Colq14}.  Indeed, it is no exaggeration to state that \textit{low and under-powered studies are the greatest threat to meta-analysis and constructing meaningful bodies of software engineering knowledge} \cite{Butt13,Szuc17}.

It is ironic that the success of software analytics research in being able to uncover patterns and effects in large, complex and noisy data sets is also its undoing.  This, coupled with the lack of guiding theory (which makes it difficult to differentiate between ``winner's curse" and some substantive effect\footnote{\RED Huang et al.~\cite{Huan17} describe a parallel situation in genomics where millions of associations can be potentially discovered but the lack of theory means there is little guidance for researchers. \BLACK}), agreed methods of analysis and excessive researcher degrees of freedom \cite{Silb15,Loke17} results in low-powered studies and a high probability of over-estimating effect sizes that are hard for other researchers to replicate \cite{Ioan05,Shep14,Jorg16}.  As an antidote, we strongly recommend:

\begin{enumerate}
\item Determining in advance the smallest effect size of interest (SESOI) \cite{Lake17}.  This may entail debate with the problem owners, typically practitioners.  It also helps address our ``Bad Smell \#1" by reducing the likelihood of conducting research that nobody cares about.
\item Also, undertake power calculations in advance to determine the likelihood of the study finding an effect of the expected size.  Where possible, when there are relevant previous studies, use a bias-corrected pooled estimate \cite{Schm14} to help address likely exaggerated estimates.
\item If the power is low either modify the study design \cite{McCl00}, use a larger sample or investigate a different question.
\end{enumerate}

\noindent
The ``smell" associated with this kind of problem can be surprisingly subtle.  However, one clue can be an argument along the lines of:
\begin{quote}
    Despite a small sample size we were able to detect a highly significant effect with a very low $p$. Thus it is concluded that the effect must be even stronger than it first appears given the inherent difficulties in detecting it.
\end{quote}

\noindent
Unfortunately, nothing can be further from the truth.  If the study is under-powered, possibly due to a small sample, then it is extremely probable that the effect is a false positive or a Type I error \cite{Loke17}.  If the study has low power then that is a problem and one that cannot be fixed by hunting for small $p$ values.
\RRED The problem that arises from small samples is that the variability will be high and only the studies that by chance obtain a sufficiently low p-value --- and therefore satisfy the acceptance criterion typically $\alpha=0.05$ --- and these will of necessity have exaggerated estimates of the effect. This is sometimes known as the ``winner's curse''.  The probability that an observed effect that reaches statistical significance actually reflects the true effect can be computed 
as $([1 - \beta] \times R) / ([1- \beta] \times R + \alpha)$ 
where $(1-\beta)$ is the power, $\beta$ is the Type II error, $\alpha$ is the Type I error and R is the pre-study odds that an investigated effect is truly non-null among all the effects being investigated) \cite{Loke17,Szuc17}. \BLACK
Note that this issue  leads to the next ``bad smell''.

\subsection{$p<0.05$ and all that!}

Though widely deprecated~\cite{carver93,Simm11,Colq14},
 null hypothesis significance testing (NHST) persists as the dominant approach to statistical analysis. Hence, the following fictitious\footnote{Since problems with the use of p-values are widely agreed to be endemic \cite{Simm11,Jorg16} we consider it invidious to single out individuals.} examples ``smell bad'':
\begin{quote}
    {\em Our analysis yields a statistically significant correlation $(r = 0.01;n = 18000;p = 0.049;\alpha = 0.05)$.}
\end{quote}

\noindent
and

\begin{quote}
    {\em Table X summarizes the 500 results of applying the ABC inferential test and reveals that 40 are statistically significant at the customary threshold of $\alpha=0.05$.}
\end{quote}

\noindent
In the first instance the low (and under the $\alpha=0.05$ acceptance threshold for Type I errors)  $p$-value should not be conflated with a practical, real world effect.   A correlation of $r=0.01$ is trivial and to all practical purposes indistinguishable from no correlation.  It is simply the artifact of the very large sample size ($n=18000$) and the logic of NHST which demands that empirical results be dichotomised into true (there is a non-zero effect) and false (the effect size is precisely zero).  

Hence, methodologists encourage the notion of the {\em smallest effect size of interest} (SESOI).  This should be specified \textit{before} the research is conducted.  It also has the merit of re-engaging the researchers with the software engineering problems they are trying to solve.  So for, say software effort prediction, after consultation with the problem owners we may conclude that the SESOI is to be able to reduce the mean inaccuracy by 25\%. Generally, although not necessarily, effect sizes are expressed as standardized measures to enable comparisons between settings so the above SESOI could be normalized by the variability (standard deviation) of the observations. For accessible overviews see \cite{Elli10,Cohe92}.  It is useful for practitioners if the article report the effect size with confidence limits.  After all, we experience effects not $p$ values!
 
In the second example of a ``bad smell'', the problem relates to the practice of performing excessive numbers of statistical tests and only attending to the ones that are ``significant".  Given that 500 tests are performed and we accept the likelihood of wrongly rejecting the null hypothesis at 1 in 20 we would expect to observe 40 ``positive" results just with random data!  A range of corrections have been proposed \cite{Bend01}.  Controlling for the false discovery rate using the Benjamini-Hochberg approach \cite{Benj01} can be appropriate as it corrects the acceptance threshold without being as severe as the traditional Bonferroni correction ($\alpha / t$ where $t$ is the number of tests being undertaken).

One way to avoid excessive statistical tests is to cluster results before performing a statistical analysis. In this approach, instead of saying ``what distributions are different?'' some grouping operator is applied prior to the application of statistics.  For example, consider the   Scott-Knott procedure recommended by Mittas and Angelis \cite{mittas2013ranking}.  This method sorts a list of $l$ treatments with $\mathit{ls}$ measurements by their median score. It then splits $l$ into sub-lists $m,n$ in order to maximize the expected value of differences in the observed performances before and after divisions.  The Scott-Knott procedure identifies one of these divisions to be `best' as follows.  For lists $l,m,n$ of size $\mathit{ls},\mathit{ms},\mathit{ns}$ where $l=m\cup n$, the `best' division maximizes $E(\Delta)$; i.e., the difference in the expected mean value before and after the split: 
 \[E(\Delta)=\frac{ms}{ls}abs(m.\mu - l.\mu)^2 + \frac{ns}{ls}abs(n.\mu - l.\mu)^2\]
Scott-Knott then checks if the `best' division is actually useful. To determine this the procedure applies some statistical hypothesis test $H$ to check if $m, n$ are significantly different. If so, Scott-Knott then recurses on each half of the `best' division.
 
For a specific example, consider the results from $l=5$ treatments:

{\small 
\begin{verbatim}
        rx1 = [0.34, 0.49, 0.51, 0.6]
        rx2 = [0.6,  0.7,  0.8,  0.9]
        rx3 = [0.15, 0.25, 0.4,  0.35]
        rx4=  [0.6,  0.7,  0.8,  0.9]
        rx5=  [0.1,  0.2,  0.3,  0.4]
\end{verbatim}
}

\noindent
After sorting and division, Scott-Knott determines:
\bi
\item Ranked \#1 is rx5 with median= 0.25
\item Ranked \#1 is rx3 with median= 0.3
\item Ranked \#2 is rx1 with median= 0.5
\item Ranked \#3 is rx2 with median= 0.75
\item Ranked \#3 is rx4 with median= 0.75
\ei
Note that Scott-Knott finds little meaningful difference between rx5 and rx3. Hence, they have the same rank, even though their medians differ.  For a larger example of this kind of analysis, see Figures~\ref{bad} \& \ref{fig:sk}.

The Scott-Knott procedure can ameliorate the excessive statistical tests problem.  To see this, consider an all-pairs hypothesis test of 10 treatments. These can be compared statistically 
\mbox{$(10^2-10)/2=45$} ways.  To substantially reduce the number of pairwise comparisons, Scott-Knott only calls on hypothesis tests {\em after} it has found splits that maximize the performance differences.  Hence, assuming binary splits of the whole data and all splits proving to be different,
Scott-Knott would be called $\log_2(10)\approx 3$ times. 
A 95\% confidence test run for these splits using Scott-Knott would have a confidence threshold of: 
\mbox{$0.95^{3} = 86$} \% assuming no corrections are made for multiple tests..
 
\begin{figure}[!t]
\begin{center}
{\scriptsize\begin{tabular}{r|rrrrrrrrrr}
treatment &  \multicolumn{9}{c}{Results from multiple runs}\\\hline
no-SWReg &   174 &    38 &     0 &    32 &    50 &   141 &    34 &   317 &   114\\
no-SLReg &   127 &    73 &    45 &    35 &    47 &   159 &    37 &   272 &   103\\
no-CART-On &   119 &   257 &   425 &   294 &   181 &    98 &   409 &   520 &    16\\
no-CART-Off &   119 &   257 &   425 &   294 &   181 &    98 &   409 &   520 &    16\\
no-PLSR &    63 &   213 &   338 &   358 &   163 &    44 &    14 &   520 &    44\\
no-PCR &    55 &   240 &   392 &   418 &   176 &    45 &    12 &   648 &    65\\
no-1NN &   119 &    66 &    81 &    82 &    70 &   110 &   118 &   122 &    92\\
log-SWReg &   109 &    23 &     9 &     7 &    52 &   100 &    12 &   164 &   132\\
log-SLReg &   108 &    61 &    49 &    53 &    55 &   156 &    28 &   183 &   101\\
log-CART-On &   119 &   257 &   425 &   294 &   181 &    98 &   409 &   520 &    16\\
log-CART-Off &   119 &   257 &   425 &   294 &   181 &    98 &   409 &   520 &    16\\
log-PLSR &   163 &     7 &    60 &    70 &     2 &   152 &    31 &   310 &   175\\
log-PCR &   212 &   180 &   145 &   155 &    70 &    68 &     1 &   553 &   293\\
log-1NN &    46 &   231 &   246 &   154 &    46 &    23 &   200 &   197 &   158\\\hline
\end{tabular}}
\end{center}

\begin{raggedright}\small{NOTE: treatments have a pre-processor (`no' =  no pre-processor; `log' = use a log transform) and a learner (e.g., `SWReg' = stepwise regression; CART =  decision tree learner that post-prunes its leaves after building the tree).}
\end{raggedright}

\caption{How not to present data (in this case, percent error estimations where the treatment in column one was applied to multiple randomly selected stratifications of the data). Note that there is (a)~no sort order of the treatments from most to least interesting;  (b)~no summary of the distributions; (c)~no indication  to indicate what values are best; and  (d)~no clustering together of results that are statistically indistinguishable. For a better way
to present this data, see Figure~\ref{fig:sk}. }\label{bad}
\end{figure}
\begin{figure}
 {\scriptsize \begin{center} 
\begin{tabular}{rrrrl}
RANK  &    TREATMENT &MEDIAN & (75-25)TH\\\hline

   1  &    no-SWReg  &    50&      107 &  \verb+ - o   --------                + \\
   1  &   log-SWReg  &    52&       97 &  \verb+   o  --                       + \\
   1   &  log-SLReg  &    61 &      55 &  \verb+  -o ----                      + \\
   1   &   log-PLSR  &    70 &     132 &  \verb+ -  o   -------                + \\
   1  &    no-SLReg  &    73  &     82 &  \verb+  - o -------                  + \\
   1   &     no-1NN  &    92  &    37  &  \verb+     o                         +\\ \rowcolor{gray} 
  2   &    log-PCR  &   155   &   142  &  \verb+ ---    o ----------------     + \\\rowcolor{gray} 
   2   &    log-1NN  &   158  &    154 &  \verb+  -     o --                   +\\ \rowcolor{gray} 
  2   &    no-PLSR  &   163   &   294 &  \verb+ --     o        --------      + \\\rowcolor{gray} 
   2  &      no-PCR   &  176   &   337 &  \verb+ --      o         ----------- + \\
   3  &log-CART-Off   &  257   &   290 &  \verb+ -----      o      ------      + \\
   3 &  log-CART-On   &  257   &   290 &  \verb+ -----      o      ------      + \\
   3 &   no-CART-On  &  257    &  290  &  \verb+ -----      o      ------      + \\
   3  & no-CART-Off   &  257   &   290 &  \verb+ -----      o      ------      +\\  
\end{tabular} 
 
 \end{center}
}
\caption{Using Scott-Knott, there is a better way to  representation  the error results see in  Figure~\ref{bad} (and  {\em lower}
error
values are {\em better}). Note that, here, it is clear that the ``SWReg'',``CART'' treatments are performing best and worst (respectively). Results from each treatment sorted by their median (and the first rows with less error are better than the last rows
with most error). Results presented numerically and visually (right-hand column marks the 10th, 30th, 50th, 70th, 90th percentile range; and `o' denotes the median value).
Results  are clustered in column one using a Scott-Knott analysis.  Rows have different ranks if Cliff's Delta reports more than a small effect difference and a bootstrap test reports significant differences.  A background gray share is used to make different ranks stand out. 
}\label{fig:sk}
\end{figure}

 
Implementations that use bootstrapping and a non-parametric effect size test (Cliff's Delta) to control its recursive bi-clustering are available as open source package, see \href{https://goo.gl/uoj3FL}{goo.gl/uoj3FL}.  
 That package converts data such as Figure~\ref{bad} into reports like Figure~\ref{fig:sk}.

\subsection{Assumptions of normality and equal variances in statistical analysis}

A potential warning sign or ``bad smell'' is when choosing statistical methods, to overlook the distribution of data and variances and focus on the ``important" matters of description e.g., using measures of location such as the mean, and the application of inferential statistics such as t-tests.

It is now widely appreciated that applying traditional parametric statistics to data that violates the assumptions of normality, i.e., they deviate strongly from a Gaussian distribution \textit{can} lead to highly misleading results \cite{Kitc02}.  However, four points may be less appreciated within the software engineering community. 
\begin{enumerate}
    \item Even minor deviations can have major impacts, particularly for small samples \cite{Wilc12}.  These problems are not always easily detected with traditional tests such as Kolmogorov-Smirnov so visualization via qq-plots should be considered.
\item Differences between variances of the samples or heteroscedasticity, may be a good deal more problematic than non-normality \cite{Erce08}. Thus the widespread belief that t-tests are robust to departures from normality when sample sizes are approximately equal and $>25$ is unsafe \cite{Erce08,Zimm98}.  
\RED \item Although different transformations, e.g., log and square root can reduce problems of asymmetry and outliers they may well be insufficient to deal with other assumption violations. \BLACK
\item Lastly, there have been major developments in the area of robust statistics over the past 30 years that are generally more powerful and preferable to non-parametric statistics.  This includes trimming and winsorizing \cite{Hoag83,Kitc17}.
\end{enumerate}

Since SE data sets can be highly skewed, heavy tailed or both, \RED we recommend performing robust tests for statistical inference. Note that we believe they should be preferred (wherever possible), to non-parametric statistics that are based on ranks. The latter, \BLACK for example the Wilcoxon signed rank procedure can lack power and are problematic in face of many ties \cite{Blai85}.  Kitchenham et al.~\cite{Kitc17} provide useful advice in the context of software engineering and Erceg-Hurn and Mirosevich an accessible general overview \cite{Erce08}.  For an extensive and detailed treatment see \cite{Wilc12}.  

Another approach is to use non-parametric bootstrapping (see Efron and Tibshirani~\cite[p220-223]{efron93}). Note that  Figure~\ref{fig:sk} was generated with such a bootstrap test.  With the advent of powerful computers, Monte Carlo and randomization techniques can be very practical alternatives to parametric methods \cite{Manl97}.  

\RED
\subsection{No Data Visualisation}\label{tion:viz}

\begin{figure}[!t]
\begin{center}
\includegraphics[width=4in]{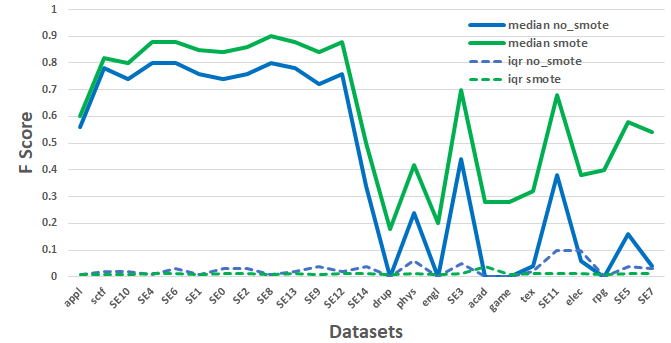}
\end{center}
\caption{\RED Effects of SMOTEing  while   text mining  25 SE data sets.  In this figure, the data sets on the x-axis are sorted by the size of the different between using and not using SMOTE.
  Medians and IQRs shown as solid and dashed lines, respectively (recall that the medians are the 50th percentile result
and the IQRs are the (75-25)th percentile). From~\cite{Krishna2016TheP}.
\BLACK}\label{fig:irqs}
\end{figure}

Whilst its customary to describe the data with a few summary statistics such as means, medians and quartiles this does not necessarily provide a check against unusual patterns and anomalies. 
Hence, we strongly advise {\em visualizing the data} first before {\em applying statistical tests}. 
For example:
\bi
\item
The horizontal box plots of  Figure~\ref{fig:sk} can  present results from a very large number of samples in a single row.  A visual inspection of that data lets an analyst check if their conclusions are refutable due to, for instance extreme outliers or high variance.
\item
For another example, consider Figure~\ref{fig:irqs} that shows results from a 5*5-cross validation experiment exploring the value of SMOTE:
\bi
\item
SMOTE is a data rebalancing technique that adjusts training
data such no class is a small minority~\cite{agrawal18}.
\item
A 5*5 cross-val experiment is an evaluation technique where $M=$ times, we randomize order of data. Each time
divide the project data into $N$ bins, train on $N-1$ bins, test on the hold out (this is repeated for all bins). This procedure produces 25 results which in Figure~\ref{fig:irqs} are reported as the medians and inter-quartile ranges (IQR).
\ei
From this visual inspection, before doing any statistics, we can observe that in no case is using SMOTE is worse that not using SMOTE. Also, as seen in the right-hand-side results of Figure~\ref{fig:irqs}, sometimes there are  very large improvements from applying SMOTE. 
\ei
\BLACK

\subsection{Not Exploring Stability}

For many reasons, software engineering data sets often exhibit large variances~\cite{menzies12}. So it is important to check if the effect you want to report actually exists, or is just due to noise. Hence:

\begin{itemize}
\item When multiple data sets are available, try many data sets.
\item Report measures of central tendency (e.g., median) \textit{and} dispersion (e.g., standard deviation or interquartile range (IQR) (75th-25th percentile).
\item Do not just average results across N data sets. Instead, summarize the \textit{spread} of results across your data sets which might include the creative use of graphics e.g. 
\RED see Figure~\ref{fig:irqs} above \BLACK. Recall
in that figure that the dashed lines at bottom
report the variability in the 25 results shown for each
data set in that figure. The key point here is that this
variability is very small; i.e.  differences between the median results is often much larger than the variability introduced by a stochastic selection of the train set (but, of course, that visual impressions requires a sanity check-- which is the role of statistical tests). Meta-analysis is extremely difficult without dispersion statistics for the results needed to construct confidence intervals \cite{Bore09,Schm14}.
\item If exploring only one data set, then repeat the analysis perhaps 20 to 30 times (and then explore the distributions generated this way with a bootstrap procedure i.e., sampling with replacement). 
\item Temporal validation: given data divided into some time series (e.g., different versions of the same project) then train on past data and test on future data. Our recommendation here is to keep the size of the future test set constant; e.g., train on software versions $K_{-2,-1}$, then test on version $K$.  Such validation enhances realism and is likely to be far more interesting to the intended users of the research.
\item Cross-project validation: train on another project data, then test on the project. Note that such cross-project transfer learning can get somewhat complex when the projects use data with different attribute names or features.
\item Within-project validation (also called ``cross-validation''): divide the project data into $N$ bins, train on $N-1$ bins, test on the hold out. Repeat for all bins.  
\end{itemize}


Note that, temporal validation (and leave-one-out cross-validation LOOCV) do not produce a range of results since the target is always fixed. However, cross- and within-validations produce distributions of results and so must be analyzed with appropriately so that both measures of central tendency (e.g., mean or median) and dispersion (e.g. variance or IQR) are reported. As a generally accepted rule, in order to obtain, say,  \RED 25 samples of cross-project learning, select
 90\% of the data, at random,
 25 times. 
 \BLACK And to obtain, say, 25 samples of within-project learning, $M=5$ times shuffle the order of the data and for each shuffle, and run $N=5$ cross-validation.  Caveat: for small data sets under, say, 60 items, use $M,N=8,3$ (since $8 \times 3 \approx 25$).

\subsection{Not Tuning}\label{tion:tune}

Many machine learners, including data miners, have poorly chosen defaults~\cite{van2017automatic, herodotou2011starfish, fu2016, tantithamthavorn2016automated, agrawal16, agrawal18}.  Numerous recent results show that the default control settings for data miners are far from optimal. For example, when performing defect prediction, various groups report that tuning can find new settings that dramatically improve the performance of the learned mode~\cite{fu2016, tantithamthavorn2016automated,tantithamthavorn2018impact}. As a specific of that tuning, consider the standard method for computing the distance between rows $x$ and $y$ (where each row contains $i$ variables):
  \[d(x,y)=\left(\sum_i\left((x_i-y_i)^2\right)\right)^{1/2}\]
 When using SMOTE to re-balance data classes, tuning found it useful
 to alter the  ``2'' value in the above equation to some other value
(often, some number as high as ``3'').

One particularly egregious ``smell" is to compare a heavily tuned new or pet approach with other benchmarks that are untuned.  Although, superficially the new approach may appear far superior what we really learn is using the new technique well can outperform using another technique badly.  This kind of finding is not particularly useful.

As to how to tune, we {\em strongly} recommend against the ``grid search'' i.e., a set of nested for-loops that try a wide range of settings for all tuneable parameters. This is a slow approach; an exploration of grid search for defect prediction and found it took days to terminate~\cite{fu2016}.  Not only that, it has been reported that grid search can miss important optimizations~\cite{Fu16}. Every grid has “gaps” between each grid division which means that a supposedly rigorous grid search can still miss important configurations~\cite{bergstra2012random}. Bergstra and Bengio~\cite{bergstra2012random} comment that for most data sets only a few of the tuning parameters really matter – which means that much of the run-time associated with grid search is actually wasted.  Worse still, Bergstra and Bengio observed that the important tunings are different for different data sets, a phenomenon that makes grid search a poor choice for configuring software analytics algorithms for new data sets.
 
Rather than using ``grid search'', we recommend very simple optimizers (such as differential evolution~\cite{Storn1997}) can suffice for finding better tunings. Such optimizers are very easy to code from scratch. They are also readily available in open-source toolkits such as jMETAL\footnote{\href{http://jmetal.sourceforge.net}{jmetal.sourceforge.net}} or DEAP\footnote{\href{http://github.com/DEAP/deap}{github.com/DEAP/deap}}.

\subsection{Not Exploring Simplicity}\label{tion:simple} 

One interesting, and humbling, aspect of using Scott-Knott tests is that, often, dozens of treatments can be grouped together into just a few ranks. For example, in the Scott-Knott results of~\cite{ghotra2015}, 32 learners were divided into only four groups. 

While not all problems succumb to simple approaches, it is wise to compare a seemingly sophisticated method against some simpler alternative. Often when we code and test the seemingly na\"{\i}ve method, we find it has some redeeming feature that makes us abandon the more complex methods~\cite{chen2018, fu2017, krishna18a}.   Moreover, the more parsimonious our models, the less prone to overfitting \cite{Hawk04}.
 
This plea, to explore simplicity, is particularly important in the current research climate that is often focused on  Deep Learning that can be a very slow method indeed~\cite{fu2017easy}. Deep Learning is inherently better for problems that have highly intricate internal structure~\cite{lecun2015deep}. Nevertheless, many SE data sets have very simpler regular structures, so very simple techniques can execute much faster and perform just as well, or better, than Deep Learning. Such faster executions are very useful since they enable easier replication of prior results.

This is not to say that Deep Learning is irrelevant for SE. Rather, it means that any Deep Learning (or for that matter any other complex technique) result should also be compared against some simpler baseline (e.g.,~\cite{fu2017}) to justify the additional complexity.  As an example, Whigham et al.~\cite{Whig15} propose a simple benchmark based on automated data transformation and OLS regression as a comparator, the idea being that if the sophisticated method cannot beat the simple method one has to question its practical value.

Three simple methods for exploring simpler options are {\em scouts}, {\em stochastic search} and {\em feature selection}. Holte~\cite{Holte1993} reports that a very simple learner called 1R can be {\em scout} ahead in a data set and report back if more complex learning will be required. More generally, by first running a very simple algorithm, it is easy to quickly obtain baseline results to glean the overall structure of the problem.

As to stochastic search, for optimization, we have often found that {\em stochastic searches} (e.g., using differential evolution~\cite{Storn1997}) performs very well for a wide range of problems~\cite{fu2016, agrawal16, agrawal18}.  Note that Holte might say that an initial stochastic search of a problem is a {\em scout} of that problem.

Also, for classification, we have found that {\em feature selection} will often find most of the attributes are redundant and can be discarded.  For example, the feature selection experiments of~\cite{menzies07} show that up to 39 of 41 static code attributes can be discarded while preserving the predictive prowess. A similar result with effort estimation, where most of the features can be safely ignored is reported in~\cite{chen2005}. This is important since models learned from fewer features are easier to read, understand, critique, and explain to business users. Also, by removing spurious features, there is less chance of researchers reporting a spurious conclusion. Further, after reducing the dimensionality of the data, then any subsequent processing is faster and easier.

Note that many articles perform a very simple linear-time feature selection such as (i)~sort by correlation, variance, or entropy; (ii)~then remove the worst remaining feature; (iii)~build a model from the surviving features; (iv)~stop if the new model is worse than the model learned before; (v)~otherwise, go back to step (ii).  Hall and Holmes report that such greedy searchers can stop prematurely and that better results are obtained by growing in parallel sets of useful features (see their 
correlation-based feature subset selection
algorithm, described in~\cite{hall2003benchmarking}).

Before moving on, we note one paradoxical aspect of simplicity research---it can be very hard to do. Before certifying that some simpler method is better than the most complex state-of-the-art alternative, it may be required to test the simpler against the complex. This can lead to the peculiar situation where weeks of CPU are required to evaluate the value of (say) some very simple stochastic method that takes just a few minutes to terminate.

\begin{table}[!t]
 \caption{The Ghotra et al.~\cite{ghotra2015} article from ICSE'15 ranked 32 different learners often used in defect prediction using a Scott-Knott procedure. Those ranks fell into four groups. The following table shows a sample of those groups.}
  \label{tbl:learners}
 \footnotesize
 \vspace{5mm}
 \begin{tabular}{l|p{1.3in}|p{2.5in}}
{\bf RANK} & {\bf LEARNER} & {\bf NOTES}\\\hline
 1 `best' & RF= \newline random forest & 
 Random forest of entropy-based decision trees.\\\cline{2-3}
 &  LR=\  Logistic regression &
 A generalized linear regression
model.\\\hline
 2 & KNN  kth-nearest &  Classify a new instance by finding $k$ examples of similar instances.
 Ghortra et al.\ suggest $K=8$.\\\cline{2-3}
 & NB=  Na{\"i}ve Bayes &  Classify a new instance by (a)~collecting mean and standard deviations of attributes in old instances of  different classes; (b)~return the class whose attributes are statistically most similar to the new instance.\\\hline
 3 & DT=   decision trees & Recursively
 divide data by selecting attribute splits
 that reduce the entropy of the class distribution.\\
 \hline
 4 `worst' & SVM= \newline support vector machines &
 Map the raw data into a higher-dimensional space where it is easier to distinguish the examples.
 \\\hline
 \end{tabular}
 \vspace{-0.2cm}
 \end{table}

\subsection{Not Justifying Choice of Learner} \label{tion:learner}

The risk, and ``bad smell", derives from the rapid growth in machine learning techniques and the ability to construct sophisticated ensembles of different learners means that the number of permutations are rapidly becoming vast.  It is trivially easy to permute an old learner $L$ into $L'$ and each new learner results in a new article.  The ``bad smell" then is to do this with no motivation, with no justification a priori as to why we might expect $L'$ to be interesting and worth exploring.

When assessing some new method, researchers need to compare their methods across a range of interesting algorithms.  
\RRED Comparing all new methods to all prior methods is a daunting task (since the space
of all prior methods is so very large). Instead, we suggest that most researchers maintain a watching brief on the analytics literature looking for prominent papers that cluster together learners that appear to have similar performance, then draw their comparison set from at least one item of each cluster.  

For learners that predict for discrete class learning, one such clustering can be seen in \BLACK Table~9 of~\cite{ghotra2015}. This is a clustering of 32 learners commonly used for defect prediction. The results clusters into four groups, best, next, next, worst
(shown in Table~\ref{tbl:learners}). We would recommend:
\begin{itemize}
\item
Selecting your range of comparison algorithms as one (selected at random) for each group;
\item 
Or selecting your comparison algorithms all from the top group.
\end{itemize}

For learners that predict for continuous classes, it is not clear what is the canonical set, however, Random Forest regression trees and logistic regression are widely used. 
 
For search-based SE problems, we recommend an experimentation technique called (you+two+next+dumb), defined as
\begin{itemize}
\item
`You' is your new method;
\item
`Two' are well-established widely-used methods (often NSGA-II and SPEA2~\cite{sayyad2013});
\item A `next' generation method e.g. MOEA/D \cite{zhang2007moea}, NSGA-III \cite{deb14};
\item and one `dumb' baseline method (random search or SWAY~\cite{chen2018}).
\end{itemize}

\section{Discussion} \label{tion:Disc}

Methodological debates are evidence of a community of researchers rechecking each other's work, testing each other's assumptions, and refining each other's methods. 
In SE we think there are too few articles that step back from specific issues and explore the more general question of ``what is a valid article?''. 

To encourage that debate, this article has proposed the idea of a ``bad smell'' as a means to use surface symptoms to highlight potential underlying problems.  To that end, we have identified over
a dozen ``smells''.

This article has argued that, each year, numerous research articles are rejected for violating the recommendations of this article. Many of those rejections are avoidable --- some with very little effort --- just by attending to the bad smells listed in this article.  This point is particularly important
for software practitioners.  That group will struggle to get their voice heard in the research literature unless they follow the established norms of SE articles.  To our shame, we note that those norms are rarely recorded –- a problem that this article is attempting to solve.

We make no claim that our  list of bad smells is exhaustive --  and this is inevitable. For example, one issue we have not considered here is assessing the implication of using different evaluation criteria; or studying the generality of methods for software analytics.  Recent studies~\cite{krishna18a} have shown that seemingly state-of-the-art techniques for tasks such as transfer learning that were designed for one domain (defect prediction) in fact translate very poorly when applied to other domains (such as code-smell detection, etc.).
Nevertheless, the overall the aim of this article is to encourage more debate on what constitutes a ``valid'' article, recognizing that validity is not a black and white concept (hence the quotation marks).  We hope that further debate, across the broader research community, significantly matures this list.

We think that such a debate could be enabled as follows:
\begin{enumerate}
    \item Read more, reflect more. Study the methods deployed in software analytics and beyond.
    When reading a article, consider not just the content of the argument, but also its form.
    \item Promote and share ``bad smells" and anti-patterns as a means of sharing poor practice but with the goal of promoting and good practice.
    \item Pre-register studies and identify the smallest effect size of (practical) interest.
    \item Analyze the likely power of a study in advance. If too low  revise the study design (e.g., collect more data or perform a within-type analysis) or change the research.
    \item Strive toward reproducible studies
    results.
     \item Develop, agree and use community standards  for software analytics studies.
     \end{enumerate}
Further to the last point, as we mentioned above, this paper is part of a strategic plan to develop community standards in software analytics.  To that end, we hope it inspires:
     \bi
     \item More  extensive discussions about best and worst paper writing practices; 
\item A workshop series where researchers meet to debate and publicise our community's views on what constitutes a ``good''  paper;
\item Subsequent articles
that extend and improve the guidance offered above.
\ei
  
\section*{Acknowledgments}
\noindent
We extremely grateful to the editor and referees for their detailed and constructive comments.

\section*{References}

\bibliographystyle{elsarticle-alpha.bst}

\BLACK
\end{document}